# Tri-modal Charge Transport in Dilute Polar Liquid based Nanoparticulate Colloidal Dispersions

Purbarun Dhar, Arvind Pattamatta and Sarit K. Das*

Department of Mechanical Engineering, Indian Institute of Technology, Madras

*Electronic mail: skdas@iitm.ac.in

Phone: +91-44-2257 4655

Fax: +91-44-2257 4650

**Abstract**

The dominant modes of charge transport in variant polar liquid based nanoparticulate colloidal dispersions (dilute) have been theorized. Three major interacting modes with independent existence, viz. Electric Double Layer formation conjugated electrophoresis, particle polarization with consequent inter-particle electrostatic interactions and coupled electro-thermal diffusion arising out of Brownian randomization, have been identified. An analytical model based on discrete interactions of the charged particle-fluid domains explains the various behavioral aspects of such dispersions, as observed and validated from detailed experimental analysis.

    The past decade has witnessed immense response from the academic community towards comprehending the enhanced transport phenomena innate to liquid based nanoparticle dispersion systems compared to the base liquids. However, the majority of initiatives have focused towards understanding the enhanced thermal [1-6] and momentum [7-10] transport in nano-dispersions.

Consequently other fundamental transport properties like electrical conductivity within nano-dispersed media have often received a blind eye.

Despite being uncharged entities, nano-particles induce excess electrical conductivity to the polar base fluids, as evident from the few experimental reports [11-14]. The enhancement in electrical conductivity is suggestive of enhanced charge transport within the fluid. Among recent endeavors to understand the electrical conductivity of such dispersions, the approaches have been purely experimental and moreover, confined to a single class of nanoparticle and base fluid combinations. Consequently, the governing parameters and mechanisms of charge transport as of yet lack complete clarity. Furthermore, for dilute dispersions (concentration $\leq$ 2 vol. %), existing theories [15-19] and observations [20-24] on the electrical properties of nano-dispersions either over or under predict the electrical conductivities due to grossly different mechanisms involved in concentrated systems. It therefore poses a necessity to comprehend the underlying physics of charge induction and transport in dilute nano-dispersions.

An essential initialization to analytical modeling lies in determining an elemental domain representing the bulk system. In the present scenario, a twin domain approach has been incorporated. A primary elemental domain representing the fundamental functional unit of the dispersed system (Fig. 1(a)) is exploited to derive the governing analytical equations. A secondary domain (Fig.1 (b)) maps all possible inter-domain interactions among the primary domains, from the point of view of continuum approach.

[FIGURE 1]

The proposed primary domain mimics concentric spheres in space and the dimensions are computed from the dispersion concentration; the validity of the approach being strengthened by the fact that the dispersions are dilute. The primary domain diameter '$a_d$' is expressed as a function of dispersed nanoparticle mean diameter '$d_{np}$' and the volume fraction of loading '$\varphi$' as

$$a_d = d_{np} \varphi^{-1/3} \qquad (1)$$

The first dominant mode theorized is the charge transport augmentation resulting from formation of Electric Double Layer (EDL) at the nanoparticle-fluid interface and the consequential electrophoresis in the presence of an electric field. Inevitably by nature, foreign systems in interfacial contact with polar fluids develop an EDL around the phase interfacial boundary, a result of the polar fluid molecules interacting with the surface atoms or stray charges distributed on the solid media; the thickness of the EDL being expressible by the Debye thickness ($\kappa^{-1}$). Evaluation reveals that even for water, the most polar liquid used in the present study, $\kappa^{-1}$ (order of ~ $10^0$ nm) is an order of magnitude smaller than '$a_d$' (order of ~ $10^1$ nm). If the ratio of the EDL thickness to the magnitude of '$a_d$' be proposed as an analogous Knudsen number ($Kn$), its magnitude remains well short of unity. Consequently, classical laws of electrostatics can be exploited within the Guoy-Chapman layer. The relatively thin EDL leads to approximation that the Stern or Helmholtz layer is superimposed onto the particle periphery. As such the Zeta potential ($\zeta$) is near – equivalent to the particle surface potential. Treating the nanoparticle as a point Coulomb charge '$q_1$' dispersed in fluid space, the magnitude of the charge developed due to the EDL is evaluated based on '$\zeta$'. Since the charges induced within the Guoy-Chapman layer and the decaying EDL are associated to the charge induced onto the nanoparticle surface, the former cannot exist independent of the latter. This allows substantial freedom to model the charged nanoparticles along with the EDL as mobile carries within the fluidic system. The charge density '$\rho_1$' developed within the primary domain due to the charged, mobile nanoparticle is derived [a] as [1]

$$q_1 = 2\pi\varepsilon_0 \kappa_{nd} d_{np} \zeta$$
$$\rho_1 = \frac{q_1}{\pi a_d^3 / 6} = \frac{12\varepsilon_0 \kappa_{nd} \zeta \varphi}{d_{np}^2} \qquad (2)$$

where, $\varepsilon_0$ is the permittivity of free space and $\kappa$ is the relative permittivity; the subscripts '$f$' and '$np$' representing the base fluid and nanoparticle respectively.

The mobility of the nanoparticles in the fluidic domain essentially conveys that the fluid molecules comprising the primary domain are updated at each time frame without compromising the structure of the domain; a direct consequence of modeling the fluid medium as a continuum. In the presence of an applied electric field, the particle-EDL system experiences an electrophoretic force, the direction of traversal being a function of the type of charge and the localized spatial orientation of the field lines. Given the nanometer dimensions of the mobile carriers, the Reynolds number associated with the electrophoretic drift of the particles is minute. This observation coupled to the presence of thin EDL for uncharged nanoparticles; Smoluchowski's theory is utilized to determine the electrophoretic mobility of the particles.

---

[a] Supporting Information Document

The drift of the particles in the presence of the field induces a directional current component within the dispersion system. The velocity associated with the electrophoretic drift is a linear function of the applied electric field strength, thereby, in compliance to Ohm's law; the current density developed within the primary domain is a linear function of the applied field strength. The constant of proportionality evolved from mathematical manipulation of the equation yields the associated effective electrical conductivity '$\sigma_1$'; expressible as [a]

$$\sigma_1 = \frac{12\varepsilon_0^2 \kappa_f \kappa_{nd} \varsigma^2 \varphi}{d_{np}^2 \eta_f} \qquad (3)$$

where, subscript '$nd$' denotes the nano-dispersion system and η is the fluid viscosity. '$k_{nd}$' has been determined from Matijevic's equation [25]. '$\sigma_1$' exhibits temperature response in inverse equivalence to the thermal response of the fluid viscosity.

The second dominant mode augments charge transport due to the dielectric behavioral aspects of solid nanoparticles dispersed in dielectric media under the influence of the electric field '$E$' and the associated inter-domain interactions resultant of particle polarization. The interactions of the electric field with the randomly distributed charged fluid molecules within the Guoy-Chapman and Stern layers promotes the formation of non-uniformity in the effective field in the immediate circumferential neighborhood of the nanoparticle. The existence of a spatio-temporally variant non-uniform electric field in the near vicinity of the nanoparticle leads to induced polarization of the particle. Although spatio-temporally fluctuating due to the innate nature of the electric field around the peripheral neighborhood, the magnitude of the dipole moment induced on the nanoparticle can be assumed to remain constant given the near-spherical geometry of the particles. Treating the nanoparticle as a solid dielectric sphere with the potential within the sphere satisfying Laplace equation, the charge density '$\rho_2$' induced onto the particle in correspondence to the generated dipole moment; as predicted by the Clausius-Mosotti theorem, is expressed as

$$\rho_2 = \frac{6\varepsilon_0 \kappa_f}{d_{np}} \left( \frac{\kappa_{np} - \kappa_f}{\kappa_{np} + 2\kappa_f} \right) E \qquad (4)$$

In the absence of an external field, the solitary mode of electrostatic interactions among any two neighboring domains exists due to the presence of the charge induced onto the particle by the EDL. The augmented charge on the nanostructure induced by the dipole formation leads to enhanced charge transport with respect to the mutual electrostatic interactions among two neighboring domains. At any given instant of time, any random particle exists in equilibrium under the mutual influence of the twin forces of electrostatic repulsion and inertia of the particle. The spatial location of a particle at a fixed frame of time is influenced by the electrostatic repulsion from the closest neighbor at the said instance. At the limit of closest approach of two particles, the electrostatic energy accumulated within the particle due to approach against repulsion is converted to kinetic energy, and the particle gains momentum in a direction opposite to that of initial approach. This culminates in a phase where the spatial location of the particle is predominantly influenced by its present state of inertia. An interplay of the two forces leads to randomized oscillatory behavior of the particles with respect to the nearest neighboring domains (Fig. S2 [a]).

To model the component of electrical conductivity borne out of the random electrostatic interactions coupled with the augmented charge in presence of the external field, an effective analytical treatment of the mean velocity of interactions needs to be performed. Considering the separation among two neighboring particles under the repulsive influence of the surrounding family of particles, the magnitude of the mean limit of separation is determined to be of the order of '$a_d$'. At this spatial separation, owing to its inertia, the nanoparticle invades into the repulsive domain of a neighboring particle, experiencing an increasing force of repulsion. At the limit of closest approach, the particle gathers inertia in the direction opposing its approach, and the near harmonic behavior cycles as it approaches a new neighbor. Therefore, the point of mean separation behaves as a stable spatial node of equilibrium for the nanoparticle. Treating the inertial acceleration as the ratio of the mean oscillatory velocity '$u_m$' about the mean position to the time period of relaxation '$\tau$', a force balance analysis of the electrostatic repulsion and inertia yields [a]

$$u_m = (4\pi^2 \varepsilon_0 \kappa_f)^{-1} \frac{6q_1 \tau}{a_d^2 d_{np}^2 \rho_{np}} \tag{5}$$

---

[a] Supporting Information Document

The expression for mean velocity manifests as a linear function of the time period of relaxation between two successive events of complete repulsion at the limits of closest approach of two particles. The relaxation time parameter is modeled as the mean time-scale the particle utilizes to traverse the distance between two successive repulsions. Therefore, '$\tau$' is expressed as the ratio of the mean velocity '$u_m$' to the mean free path '$\lambda$' i.e. of the order of the mean limit of separation, ~ '$2a_d$'. The interaction velocity has been modeled on the basis of '$q_1$'; the charge induced due to the EDL, since it is the innate source of charge in the primary domain. The charge induced due to polarization exists only in the presence of the external field and the augmented charge begins interactions with the velocity '$u_m$'. On the basis of the order-of-magnitude analysis of the time scale of interactions '$\tau$' and the maximum limit of separation, the mean velocity of interactions is expressible as [a]

$$u_m = \frac{\varsigma}{d_{np}} \sqrt{\frac{12\varepsilon_0 \kappa_f \varphi^{1/3}}{\rho_{np}}} \tag{6}$$

where '$\rho$' denotes density. The polarized particles interact with its neighbors with the velocity '$u_m$', thereby; it is analogous to the drift velocity of the charge carriers. The polarized particles drifting with this mean velocity induces a current component within the dispersed system, which when analyzed in accordance to Ohm's law, yields an expression for electrical conductivity as [a]

$$\sigma_2 = \frac{6\varepsilon_0 \kappa_f \varsigma}{d_{np}^2} \left( \frac{\kappa_{np} - \kappa_f}{\kappa_{np} + 2\kappa_f} \right) \sqrt{\frac{12\varepsilon_0 \kappa_f \varphi^{1/3}}{\rho_{np}}} \tag{7}$$

The conductivity component '$\sigma_2$' is independent of the component '$\sigma_1$', and is a consequence of a grossly different mechanism of charge transport. As evident, it is an independent or very weak function of temperature.

The third dominant mode of charge transport exists due to the coupled thermal and electrical interactions among the fluid molecules and the charged nanoparticle. A dispersed nanoparticle in polar fluid medium experiences the formation of an EDL shell and randomized Brownian disturbance.

Fundamentally, the phenomenon can be looked upon as transport of a charged entity. Consequently, electro-thermal coupling leads to the existence of a probable weak current within such dispersions. In presence of an external field, the interaction parameters get redistributed, leading to an electro-thermal component of charge transport. A further categorization of two probable sub-modes of transport is possible. One sub-mode exists due to the coupling of the innate thermal energy of the nanoparticles and the induced polarization in presence of the external field. On the contrary, the other sub-mode arises from [3]the interactions of the innate charge on the particles owing to the presence of the EDL with the electro-thermal diffusive component introduced by the external field. The sub-component of conductivity arising due to the first sub-mode is expressed as [a]

$$\sigma_{3(1)} = \rho_{dipole} \times v_B = \frac{4\varepsilon_0 k_B T}{\pi \eta d_{np}^3}\left(\frac{\kappa_{np} - \kappa_f}{\kappa_{np} + 2\kappa_{np}}\right) \quad (8)$$

where, '$v_B$' represents the Stokes- Einstein Brownian velocity. The sub-component due to the second mode is expressed as

$$\sigma_{3(2)} = \rho_1 \times \mu_{electrothermal} = \frac{16\varepsilon_0^2 \kappa_f \kappa_d \zeta^2}{\eta d_p^2} \quad (9)$$

where, '$\mu_{electrothermal}$' represents the coupled electro-thermal diffusion coefficient.

The net electrical conductivity rendered to the dispersion due to electro-thermal coupled transport would therefore be a cumulative effect of the twin modes. However, an order of magnitude analysis for nanoparticle dispersion systems reveals that the order of '$\sigma_{3(1)}$' is at the very least ~ 100 orders of magnitude smaller than '$\sigma_{3(2)}$' for a large range of absolute temperature and wide range of base fluid relative permittivity. Hence, the contribution of the former can be accurately approximated to be dominantly over-shadowed by the latter. Hence, the transport component '$\sigma_{3(2)}$' in itself becomes equivalent to the net transport of charge due to electro-thermal interactivities.

As with majority of multi-component dynamical systems wherein the behavioral aspects of the system as a bulk entity is governed by the interactions among the fundamental structural and/or functional units of the system, the effective charge transport through nano-dispersions is governed by the interactivity among the three dominant (and independent) modes of transport discussed. The effective electrical conductivity of the dispersion is a function of the inter-modal interactions (represented by coefficients '*a*', '*b*' and '*c*') and is theorized to be the root mean square of the effective contributions of the triple modes; expressible as

$$\sigma_d = \sigma_f + \sigma_{rms}$$
$$\sigma_{rms} = \sqrt{(a\sigma_d)^2 + (b\sigma_d)^2 + (c\sigma_d)^2}$$ (10)

---

[a] Supporting Information Document

Since it is cumbersome to analytically model the interactions among individual primary domains; a simplified yet accurate method of mapping the most prominent interactions and identifying the associated odds in favor or against an event has been proposed. To achieve the relationship correlating the effective electrical conductivity of the dispersion to the tri-modal conductivities and their interactivities, a secondary domain, as illustrated in Fig. 1(b) is necessary. The illustration provides an exploded view of the primary domain (at the center of the cubic structure) and the immediate possible neighboring sister domains. The major objective of the hypothetical secondary domain lies in establishing a discretized version of the actual physical interactions, which could be infinite in number and at the same time redundant and of negligible contributions. The domain thereby serves a twin-fold purpose; analysis of interactions in the physical space to an equivalent discrete space and eliminating the redundancies that creep up while modeling the interactions in the physical space. The secondary domain distills the otherwise critical analysis; without compromising the accuracy involved. From a geometric standpoint, it is seen that a packed cubic structure, with the domain of interest at the center and surrounded by its 26 immediate neighbors is an optimally efficient discrete system to map all possible short and long range interactions. A spherical secondary domain will shadow out many of the constituent primary domains from the influence of the central domain and add redundancy. Other forms of packing structure of the primary domains will lead to possible exclusion of important neighboring entities. The interactivities in a random horizontal plane have been qualitatively illustrated in Fig. 1(c).

Evident from the illustration in Fig. 1(b), the central domain when mapped with respect to the direction of electrophoresis along the direction of the applied field; experiences electro-thermal interactions from all the 26 neighbors as well as additional electrostatic interactions from the 6 closest neighbors. Since the electrostatic interactions between two nanoparticles have a maximum separation limit of ~ '$2a_d$', the interactions are only applicable for particles whose domains are in direct contact with the domain of interest. However, electro-thermal interactions; essentially a consequence of coupled Brownian randomness and electrostatic behavior, exist between the central domain and all its immediate neighbors. Since the intermolecular separation among fluid molecules are of the order of ~ 5 Å, the uninterrupted diffusive traversal of the nanoparticle essentially dies out within the confines of the secondary domain itself, allowing for the exclusion of any interaction parameter from domains outside the cubic structure. The analysis yields that associated with the conductivity component '$\sigma_1$' there are 26 equivalent electro-thermal interactions and 6 electrostatic interactions corresponding to each of the former. Therefore, the central domain experiences a total of 26 x 6 interactions at any frozen instant of time while streaming through the fluid due to the existence of the electric field. In essence, the magnitude of the coefficient '$a$' is therefore ~ 156.

The electrostatic interactions, being dominant only along 6 discrete directions, dictate the motion of the polarized particles in presence of the external field. Since Ohm's law states that the current density vector is aligned along the electric field vector, the component of electrical conductivity '$\sigma_2$' along the field vector must be determined. Since the separation limits are equal along all the 6 cardinal directions and the primary domains are identical, it can be assumed that the probability that the direction of dielectrophoresis of the particle is dominant along the field vector is ~*1/6*. The magnitude of the weight coefficient '$b$' therefore is ~*1/6*. Similarly, the electro-thermal interactions manifest prominently along all the discrete directions possible within the domain. Thereby, the probability that the random electro-thermal diffusion occurs along the direction of the field vector is ~ *1/26;* the magnitude of the coefficient '$c$'. The thermal diffusion of the particles increases with increasing temperature. This increases the probability of interaction with particles beyond the cubic domain. Consequently, the probability of the diffusive vector being collinear to the field vector decreases. Analysis of extensive experimental data reveals that the coefficient '$c$' can be remodeled effectively to include the temperature response of the dispersed system. The expression is theorized as *$c \approx (T - T_{fp})^{-1}$*, where $T_{fp}$ denotes the freezing temperature of the base fluid. It can be verified that for aqueous systems at room temperature, the magnitude of '*c*' approaches ~ *1/26*, as predicted from the former analysis.

In specialized cases, the magnitude of the constants might not be deduced accurately without a rigorous analysis. Respecting all such possibilities, it can be generalized that the orders of magnitude of the constants '*a*', '*b*' and '*c*' will always be ~ *$10^2$*, *~$10^{-1}$* and *~$10^1$* respectively. However, the experimental data corresponding to the present work and [12, 14] have been found to exhibit close resemblance to the values predicted from the analytical model based on the theoretical magnitudes of the coefficients discussed. It has been observed that transport in both metallic and ceramic nanoparticles in variant polar and weakly polar fluids with respect to temperature and concentrations can be predicted accurately (within ±10%) utilizing the theoretical coefficients. The validation of the analytical model against experimental data has been illustrated in Figs. 2, 3 and 4.

[FIGURE 2]

The higher magnitudes of particle mobility induced surface charge density and electro-thermal diffusivity with decreasing particle dimensions leads to increased augmentation in charge transport (Fig. 2). It is noteworthy that the temperature response almost disappears in very dilute systems due to absence of additional directionality of interactions among the already far spaced neighbors. It is evident from Fig. 2 that while the magnitude of conductivity is much higher in case of smaller particles, the thermal response of the systems are nearly equivalent. This is indicative of the dominant role of liquid viscosity in the phenomenon, indirectly implying that electro-kinetic motion of the particles is a major reason behind charge transport. Although the size of Copper oxide nanoparticles (Fig 2(c)) utilized is almost the mean of the Aluminum oxide particles (Fig 2(a) and (b)), it is not directly reflected in the conductivity data due to the non-linear dependence of charge transport on particle size, as well as due to the unequal dielectric constants of the two class of particles. The importance of particle dielectric constant in comprehending the polarization governed transport is evident from Fig. 3. The marginally higher dielectric constant of Aluminum over Copper provides marginally increased values of conductivity despite the equal sizes and same base liquid. Although the second mode plays a decisive role, the equal sizes however, lead to similar thermal responses, which is again indicative of electrophoresis and electro-thermal diffusion.

**[FIGURE 3]**

The predictability of the model for weakly polar as well as high viscosity fluids such as Ethylene Glycol (EG) and Glycerol has been illustrated in Fig. 4. Water, a polar solvent with low viscosity, experiences higher augmentation in charge transport. It is evident from Fig. 4(a) that the combined effects of high viscosity and low polarity of EG compared to water results in lower augmentation in transport. The dominance of polarization induced transport over electrophoresis is observed in Fig. 4(b), wherein the enhancement in electrical conductivity for Glycerol is marginally more than that of EG, in spite of the former being more viscous. The higher value of dielectric constant of Glycerol leads to greater polarization induced transport, thereby overshadowing the viscous effects. Evident from the illustration, the trend of transport parameter with concentration, as reported in experiments [11, 13] is well mimicked by the present model; wherein very dilute systems exhibit highly non-linear behavior with concentration, whereas a more linearized response is obtained at higher concentrations. At very dilute proportions, addition of a minute population of particles leads to enhanced electrostatic and electro-thermal interactions, all existing in a relatively free environment. As the population of particles increases, the dominance of enhanced transport is fractionally over-shadowed due to increased cohesion among the particles and effects such as particle aggregation and localized crowding that creep in at higher concentrations; thereby hindering independence of the modes and freedom in transport.

**[FIGURE 4]**

The scope of the analytical model however, is not restricted within the boundaries of simply predicting the absolute magnitude of electrical conductivity of such dispersed system. The study also sheds light onto the effective role of each of the modes at different concentrations or temperatures. Evident from Fig. 5(a), the modes arising due to polarization and electro-thermal diffusion have substantial contributions at very dilute concentrations, whereas, the non-linear growth of the electrophoretic mode begins domination above a certain concentration range. This is valid for all particles having similar diametric magnitudes. The contribution of each mode with temperature (Fig. 5(b)) exhibits a different trend. While the polarization and electro-thermal diffusion contributions remain fairly constant, the electrophoretic mode shows a near-linear variance. Thereby, it can be commented that the effective non-linear trend in electrical conductivity of such systems is more dominated by the near-linear patterns in polarization in co-existence with the non-linear variation in electrophoresis.

[FIGURE 5]

To infer, the analytical model theorized successfully predicts the charge transport in dilute nano-dispersions, simultaneously capturing accurately the effects of physical parameters viz. temperature, particle size, concentration and relative permittivity and fluid viscosity and polarity. The three modes provide a clear picture of the mechanistic behavior of charge induction, interaction and transport in the presence of an electric field and involve non – adjustable interaction coefficients to map all possible inter – particle interactions important in charge transport; while predicting the decisive roles played by the modes at different conditions. Devoid of the concepts of aggregation – disaggregation, particle clustering, etc. innate to concentrated systems, the model neither over nor under-predicts the electrical conductivity of dilute dispersions. Insight into the dominance and behavior of mechanisms depending on material properties is also obtained from analysis.

**<u>Figures:</u>**

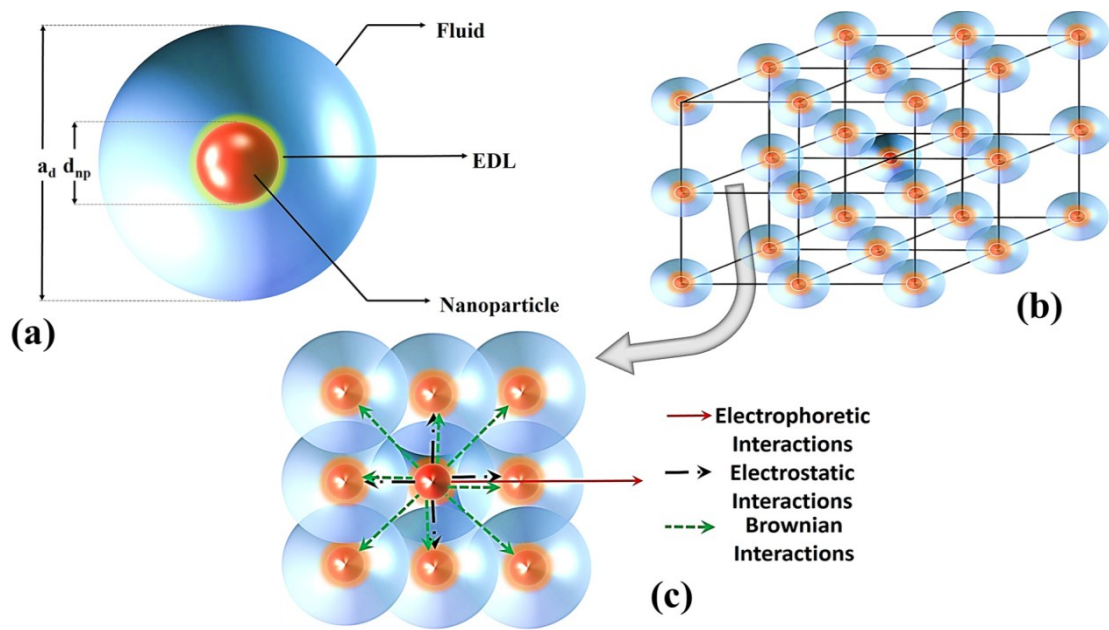

**Figure 1**: The (a) Primary cell domain (b) Neighborhood interaction mapping (secondary) domain (c) Discretized interactions of the central primary domain in (b) with its sister domains in a horizontal plane (indicated by arrow).

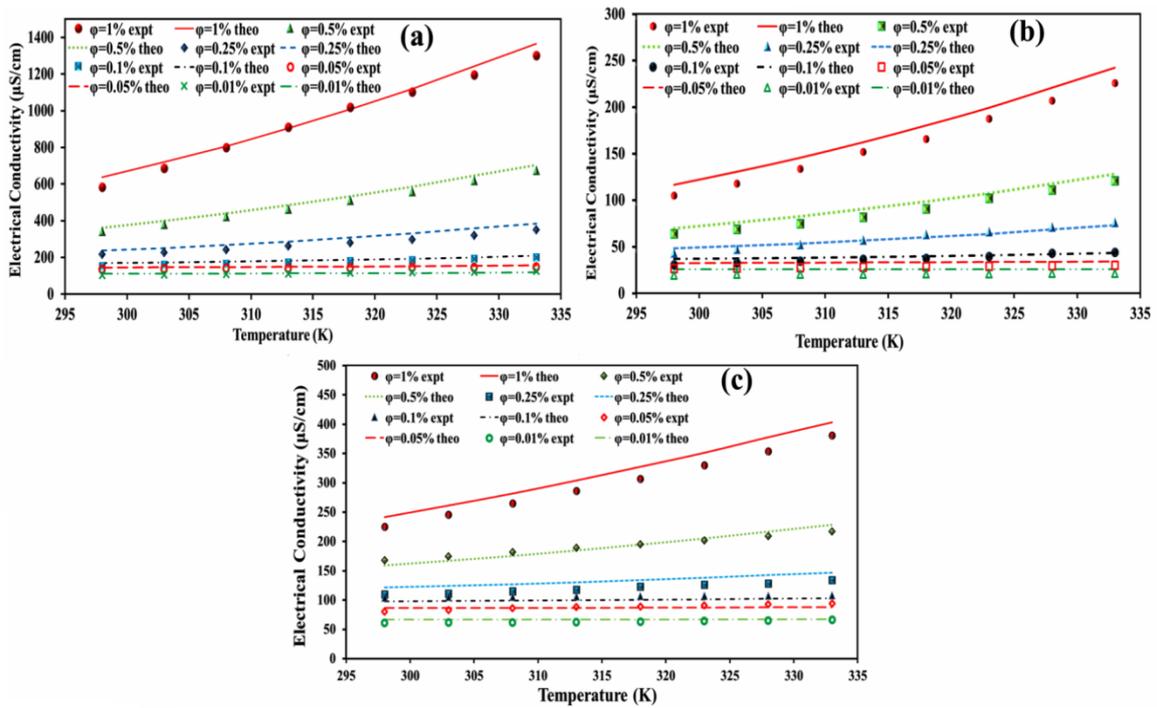

**Figure 2:** Effect of particle size and dielectric constant of nano-particle material on charge transport in water based **(a)** Alumina (20 nm) **(b)** Alumina (45 nm) and **(c)** Copper (II) Oxide (30 nm) dispersions.

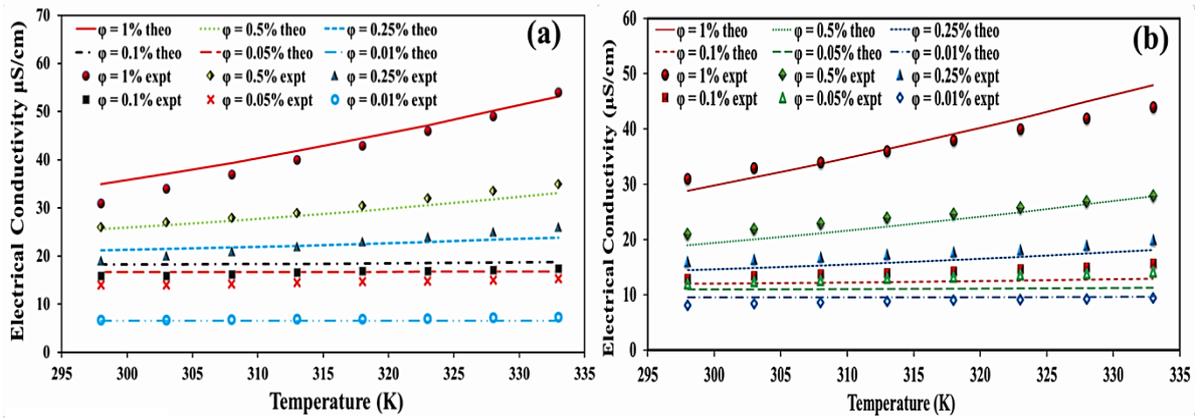

**Figure 3:** Effect of the dielectric properties of the nanoparticle on charge transport in water based **(a)** Aluminum (80 nm) and **(b)** Copper (80 nm) dispersions.

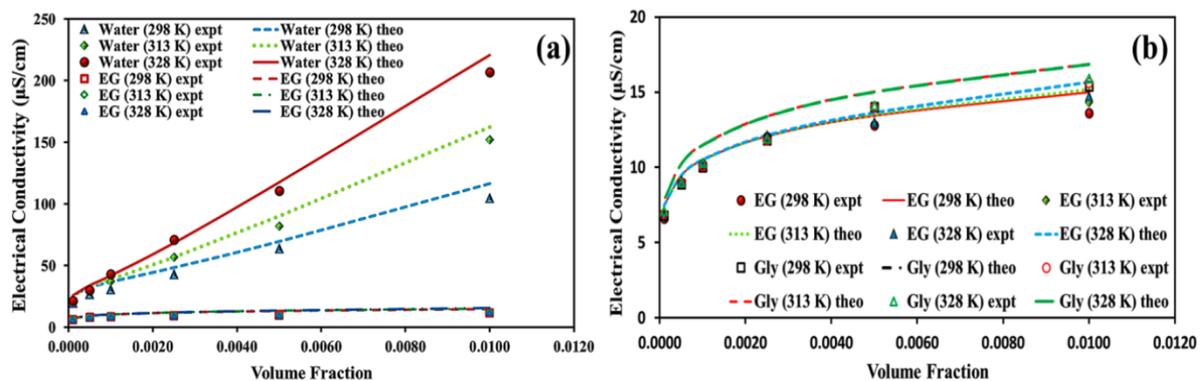

**Figure 4: (a)** Combined effects of fluid viscosity and dielectric constant on the transport of charge. **(b)** The model also accurately predicts the non-linear behavior of electrical transport with concentration; similar to experimental reports [12, 14].

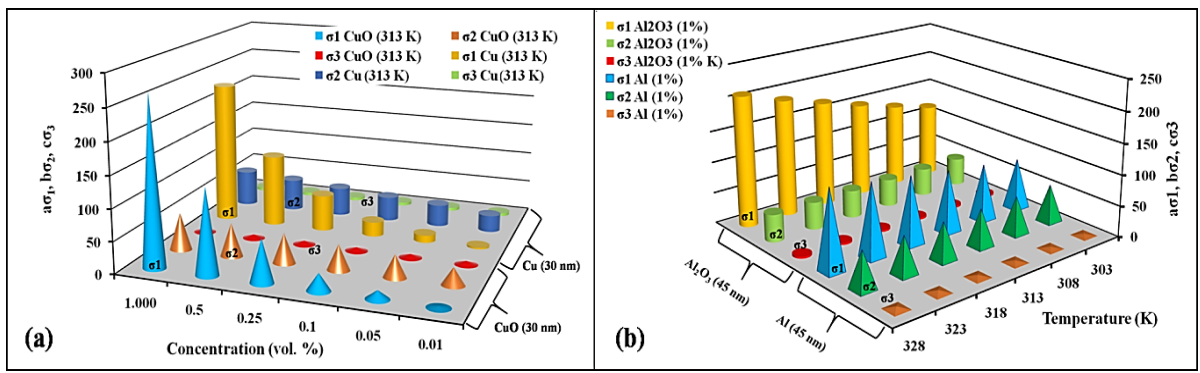

**Figure 5:** Variation of trends of the three modes with their weighted interaction coefficients (aσ$_1$, bσ$_2$, cσ$_3$) with **(a)** concentration at a constant temperature and **(b)** temperature at a constant concentration for aqueous dispersions of metallic and ceramic nanoparticles.